\documentclass[a4paper,11pt]{article}
\usepackage{aaskaiid}
\usepackage{xcolor}
\usepackage{array}
\usepackage{orcidlink}

\title{Exploring Gravitational Wave Science Frontiers with the SKAO}
\ShortTitle{Gravitational Wave Science with the SKAO}

\author[1,2,3]{Nicola Bellomo \orcidlink{0000-0002-4375-705X}}
\ShortName{Bellomo and Cole} 
\author[4]{Philippa Cole \orcidlink{0000-0001-6045-6358}}

\affiliation[1]{Dipartimento di Fisica e Astronomia ``G. Galilei'', Universit\`a degli Studi di Padova, via Marzolo 8, I-35131 Padova, Italy}
\affiliation[2]{INFN, Sezione di Padova, Via Marzolo 8, I-35131, Padova, Italy}
\affiliation[3]{INAF - Osservatorio Astronomico di Padova, Vicolo dell'Osservatorio 5, I-35122 Padova, Italy}
\emailAdd{nicola.bellomo@unipd.it}
\affiliation[4]{Department of Physics \& Astronomy, Queen Mary University of London, Mile End Road, London E1 4NS, UK }
\emailAdd{p.s.cole@qmul.ac.uk}

\abstract{
The Square Kilometre Array Observatory (SKAO) will be an important component of the global gravitational wave network.
This article provides an overview of chapter eight of the Advancing Astrophysics with the SKA II (AASKAII) book, in which gravitational waves are a new addition, since the previous edition preceded the announcement of the first detection of gravitational waves in 2016. 
The chapter investigates the impact that this new observatory will have on numerous gravitational wave science cases. 
From testing General Relativity, to measuring the properties of the nanohertz gravitational wave background and exploiting new synergies with other upcoming experiments, the SKAO will play a key role in the next decades of gravitational wave science.
}

\begin{document}
\maketitle

\section{Introduction}
Gravitational waves (GWs) were detected for the first time in 2015 from the coalescence of two black holes of stellar-origin. 
Since then, more than 300 events have been detected with the LIGO-Virgo-KAGRA network, and many thousands of events are expected to be observed with the next generation of ground-based gravitational wave detectors currently under consideration including Einstein Telescope and Cosmic Explorer. 
The Laser Interferometer Space Antenna (LISA), the first space-based gravitational wave detector, is set to launch in the mid to late 2030s and will observe gravitational wave signals from an entirely new frequency range, with supermassive black hole binary mergers and extreme mass ratio inspirals as key targets. Furthermore, in 2023, the International Pulsar Timing Array (IPTA) announced evidence for a signal consistent with a stochastic gravitational wave background within their datasets at nanoHertz frequencies. The leading explanation for this stochastic signal is the superposition of gravitational waves being emitted by the inspiral and merger of supermassive black hole binaries. These detections have driven an explosion of progress and ideas in the field, opening a new channel with which to observe the universe. 

The Square Kilometre Array Observatory (SKAO) will play a key role in this field, which has blossomed since the publication of the first SKAO Red Book ten years ago. Alongside its capabilities as a powerful pulsar timing array (PTA) network, we outline some of the ways in which SKAO observations will be harnessed to extract new gravitational wave signals and cross-correlate with datasets from other detectors. Fruitful directions include searching for new physics signals such as signatures of beyond General Relativity effects and cosmological phase transitions, synergies between SKAO pulsar timing array observations, other gravitational wave detectors and astrometric data, as well as using SKAO observations to understand large-scale structure, clustering and anisotropies of gravitational wave sources.

%%%%%%%%%%%%%%%%%%%%%%%%%%%%%%%%%%%%%%%%%%%%%%%%%%%%%%%%%%%%%%%%%%%%%%%%%%%%%%%%%%%%%%%%%%%%%%%%%%%%%%%%%%%%%%%%%%%%%%%%%%%%%%%%%%%%%%%%%%%%%%%%%%

\section{Tests of General Relativity}

The unprecedented pulsar timing residual precision achievable with SKAO will provide the opportunity to test General Relativity in novel ways. 
In~\cite{Besancon01.2026.SKA}, the authors show how the expected sensitivity of this observatory can provide the first detection of memory effects.
The GW memory effect is the permanent change in the relative position of free-falling test masses generated by the passage of a GW.
Its existence has been predicted in different forms for more than 50 years; however, it has never been experimentally demonstrated.
After a brief summary of the existing and future efforts in searching for memory effects in other GW frequency bands, namely mHz and Hz-kHz, the authors describe which sources and expected signals can be constrained with PTA measurements.
Although still in the early days of memory effect searches, the projected sensitivity of SKAO suggests that it will be possible to measure with a high-degree of confidence the effect of a GW memory burst generated by a binary of supermassive black holes (SMBHs) with masses~$M_\mathrm{SMBH} \gtrsim 10^8\ M_\odot$ at cosmological distances, providing an additional confirmation of the robustness of General Relativity.

%%%%%%%%%%%%%%%%%%%%%%%%%%%%%%%%%%%%%%%%%%%%%%%%%%%%%%%%%%%%%%%%%%%%%%%%%%%%%%%%%%%%%%%%%%%%%%%%%%%%%%%%%%%%%%%%%%%%%%%%%%%%%%%%%%%%%%%%%%%%%%%%%%

\section{Cosmological gravitational wave backgrounds}

Establishing whether supermassive black hole binaries (SMBHBs) are the main sources that contribute to the nHz GW background (GWB), or whether there exists a cosmological component, remains a vibrant and important line of research.
The discovery of even a subdominant cosmological component will have significant repercussions for our understanding of Early Universe processes, especially at energies beyond the reach of laboratory experiments.
In this sense, the astounding pulsar timing precision that will be reached with the SKAO will certainly help in further characterizing the properties of the background, such as its spectral shape and statistics.

In~\cite{Pasechnik01.2026.SKA}, the authors review the current theoretical landscape behind a very popular class of cosmological sources for the nHz background: first order phase transitions (FOPTs) occurring at temperatures around the quantum chromodynamics (QCD) scale.
FOPTs are initiated by the decay of a field trapped in a metastable vacuum into its stable vacuum configuration, and GWs are subsequently created as a result of bubble nucleation and collision, long-lived sound waves in the plasma, and magnetohydrodynamic turbulence.
Additionally, FOPTs naturally appear in many well-motivated beyond-Standard-Model scenarios that aim to address other existing open problems, such as the origin of neutrino masses or the nature of dark matter.
In particular, the contribution presents an in-depth discussion of the GW signal, its microphysics origin, and its detectability prospects associated with two promising dark sector models: supercooled phase transitions created by a dark sector with MeV/GeV symmetry breaking scale, and FOPTs generated by strongly-interacting dark sectors associated with composite dark matter.
Finally, the authors discuss FOPTs as a possible source of primordial black holes and cosmic defects, and the possibility that the GW signatures associated with the latter phenomena coexist or even dominate the FOPT GW signal itself.
In summary, the SKAO has unprecedented potential to provide a unique and complementary handle on constraining all of these scenarios.

An alternative, yet compelling cosmological background candidate is explored in~\cite{Ragavendra01.2026.SKA}: scalar-induced GWs (SIGWs).
SIGWs are primordial GWs generated by the coupling of scalar and tensor perturbations at second-order in perturbation theory; therefore, they are guaranteed to exist, especially in light of the overwhelming cosmological evidence for the existence of primordial scalar fluctuations.
Numerous inflationary models predict the existence of large, almost non-linear, scalar perturbations at small-scales: if such perturbations are present, a potentially detectable GW background is generated.
The authors of this contribution show that, given the sensitivity of SKAO, it will not only be feasible to detect such a signal but also to characterise it to high accuracy and to potentially find clear signatures of primordial non-Gaussianity.
Additionally, it is possible to provide additional constraints on the existence of parity violation phenomena in the Early Universe via measurements of the GWB's chirality. 
Measuring the properties of the GWB with SKAO will therefore enable us to trace back its characteristics to the dynamics of the Early Universe and therefore probe physics that is otherwise difficult to probe directly.

%%%%%%%%%%%%%%%%%%%%%%%%%%%%%%%%%%%%%%%%%%%%%%%%%%%%%%%%%%%%%%%%%%%%%%%%%%%%%%%%%%%%%%%%%%%%%%%%%%%%%%%%%%%%%%%%%%%%%%%%%%%%%%%%%%%%%%%%%%%%%%%%%%

\section{Synergies between detectors}

SKAO will play an important role in the emerging multi-band and multi-messenger exploration of the gravitational-wave Universe at low frequencies. In particular, SKAO observations will enable the exploration of previously un-probed frequency regimes and provide complementary observables that can be cross-correlated with both other gravitational wave detectors and electromagnetic facilities. Below three contributions to this Science Book from the gravitational wave working group, where these synergies are especially promising, are highlighted.

Binary millisecond pulsars provide a unique opportunity to probe gravitational waves in the microhertz band through precise measurements of their orbital dynamics, as proposed by \cite{Jenkins01.2026.SKA}. Gravitational waves passing through a binary system induce small perturbations in its orbital parameters which can be measured by SKAO with unprecedented precision. The orbital period of the system determines the gravitational wave frequency range to which it is most sensitive, allowing binaries with different orbital periods to act as detectors at different frequencies. Systems with significant orbital eccentricity are particularly valuable, as their orbital motion sweeps through a range of binary separations and hence generates broadband sensitivity.

The target frequency range lies between the nanohertz band probed by PTAs and the millihertz band targeted by space-based interferometers such as LISA, and is currently unexplored. Forecasts indicate that long-term SKAO observations of a handful of suitable binaries could reach sensitivities sufficient to detect a plausible stochastic background produced by SMBHBs. Because many of the same systems will also be monitored by PTA experiments, this approach naturally complements conventional pulsar timing analyses and can strengthen the multi-band detection campaign.

High-precision astrometry offers another independent method for detecting gravitational waves, as proposed by \cite{Perna01.2026.SKA}. As a gravitational wave passes through space-time, it induces small deflections in the apparent angular positions of distant sources such as stars or asteroids. Large astrometric surveys can therefore search for correlated patterns in the apparent motions of many objects across the sky.

This observable is closely related to the timing residuals measured in pulsar timing arrays, since both effects are produced by the same space-time perturbations caused by passing gravitational waves. Combining these two measurements provides a powerful opportunity for cross-correlation studies. In particular, correlated signatures between PTA timing residuals and astrometric deflections could strengthen the statistical significance of a stochastic gravitational wave background detection.

The angular deflection field induced by a stochastic gravitational wave background can be decomposed into modes on the celestial sphere. A joint analysis of astrometric and PTA data enables the reconstruction of this angular power spectrum and provides access to additional observables. Notably, correlations between the two parity modes in the angular decomposition would signal parity-violating phenomena in the early Universe.

Forecasts suggest that surveys with sensitivities comparable to those of missions such as Gaia or the Roman Space Telescope could detect the stochastic gravitational wave background with signal-to-noise ratios exceeding unity out to angular multipoles of order $l\sim10$. Cross-correlation with PTA measurements obtained using the SKAO would significantly enhance the scientific return of both fields.

Massive black hole binaries embedded in gas are among the most promising sources for joint gravitational-wave and electromagnetic observations, as laid out in \cite{Capelo01.2026.SKA}. Depending on their masses and evolutionary stage, these systems can emit gravitational waves in multiple frequency bands simultaneously while also producing electromagnetic radiation across the spectrum.

In particular, gas-embedded massive black hole binaries are expected to emit gravitational waves in the nanohertz band, accessible to PTA experiments including SKAO, or in the millihertz band targeted by LISA, depending on their mass. Furthermore, for those more massive that are detectable in the nanohertz band, gas torques from the surrounding medium can excite additional gravitational wave harmonics at higher millihertz frequencies, potentially allowing the same system to be observed across multiple gravitational wave frequency bands with both SKAO and LISA.

In addition, electromagnetic emission from these binaries may occur before, during, or after the gravitational wave-driven inspiral and merger, driven by the formation of mini-discs around the two massive black holes. The resulting electromagnetic waves cover the entire spectrum, from X-ray to radio, with the latter being detectable by SKAO in the frequency range 50–15400 MHz.

The combination of gravitational wave measurements from PTAs and LISA with SKAO radio observations opens the possibility of simultaneous GW–GW (nanohertz–millihertz) and GW–electromagnetic detections of the same systems. Such multi-band and multi-messenger observations, made possible by SKAO and LISA operating at the same time, would provide a unprecedented view of the astrophysical environment of massive black hole binaries, enabling detailed studies of their dynamical evolution, gas interactions, and the physical processes driving their electromagnetic emission.

%%%%%%%%%%%%%%%%%%%%%%%%%%%%%%%%%%%%%%%%%%%%%%%%%%%%%%%%%%%%%%%%%%%%%%%%%%%%%%%%%%%%%%%%%%%%%%%%%%%%%%%%%%%%%%%%%%%%%%%%%%%%%%%%%%%%%%%%%%%%%%%%%%

\section{Anisotropy of gravitational wave backgrounds}

Whether astrophysical or cosmological in nature, the GW background is expected to present some degree of intrinsic anisotropy connected to the nature of the sources that generated it.
The different kinds of anisotropies are generally divided into four different classes, from the most to the least dominant: loud sources, kinematic, astrophysical, and cosmological.
Loud sources and astrophysical anisotropies are associated with astrophysical backgrounds generated by SMBHBs.
In particular, the former are associated with the closest and most strongly emitting binaries, whereas the latter are due to the fact that SMBHBs trace the galaxy populations; thus, they inherit their intrinsically anisotropic distribution.
Kinematic anisotropies are common to both astrophysical and cosmological backgrounds since their origin is the observer's motion with respect to the cosmic rest frame.
Finally, cosmological anisotropies are generated by GW propagation across a perturbed Universe, as well as with the background formation mechanism at early times.

In~\cite{Cruz01.2026.SKA}, the authors investigate the detectability prospects of kinematic anisotropies - in the form of a kinematic dipole - for the AA4 and beyond configurations of SKAO.
This line of research is of primary interest since numerous large-scale structure measurements report a tension both in terms of amplitude and direction of the dipole with respect to the values inferred from the Cosmic Microwave Background (CMB).
According to the authors' forecasts, SKAO has the potential of tightening current upper limits by a factor 4-to-10, depending on the chosen configuration.
Additionally, astrometry measurements are expected to further tighten SKAO-only bounds, although the improvement is expected to be marginal.
Although this sensitivity level will not be sufficient to improve upon the accuracy of CMB measurements, it could still be used to test the robustness of the dipole amplitude and direction derived from large-scale structure measurements.
Finally, this test will also provide valuable constraints on Early Universe models that imprint a directional structure on the background itself.

Although current PTA measurements are still limited by a limited number of monitored pulsars, timing residual sensitivity, and non-uniform coverage of the sky, the SKAO will dramatically change this landscape.
As discussed in~\cite{Ragavendra01.2026.SKA}, such remarkable improvement can potentially lead to the first detection of astrophysical anisotropies of the GWB in cross-correlation with ongoing and future galaxy surveys, such as Vera Rubin, DESI, and Euclid.
This detection will confirm the nature of the GWB, and it will be eventually used to characterise the tight connection between SMBHBs and galaxy formation and evolution.

Finally, the SKAO also has the potential of working in synergy with future GW observatories in the Hz-kHz band, namely Einstein Telescope and Cosmic Explorer, thanks to its SKA-Mid neutral-hydrogen intensity mapping experiment and radio-continuum galaxy survey.
In~\cite{Bosi01.2026.SKA}, the authors highlight the potential of measuring the cross-correlation signal between the gravitational wave anisotropic distribution and probes of the large-scale structure of the Universe.
In particular, it is shown that it will be possible to reach a precision of the order of ten percent in the measurement of the bias parameter, i.e., the main parameter that dictates the strength of the clustering signal, over a large span of cosmic epochs.
Since the bias parameter naturally encodes information regarding the physics of BBH formation, the study of GW clustering will be able to shed light on the origin of black holes.
In particular, we will be able to probe whether the BHs are astrophysical or primordial in nature and to provide a first independent measurement of the binary time-delay distribution, potentially ruling out some astrophysical formation channels.

%%%%%%%%%%%%%%%%%%%%%%%%%%%%%%%%%%%%%%%%%%%%%%%%%%%%%%%%%%%%%%%%%%%%%%%%%%%%%%%%%%%%%%%%%%%%%%%%%%%%%%%%%%%%%%%%%%%%%%%%%%%%%%%%%%%%%%%%%%%%%%%%%%

\section{Conclusions and outlook}
SKAO observations will push key gravitational wave science frontiers, enabling tests of General Relativity through the detection of gravitational wave memory effects, providing unprecedented sensitivity to cosmological gravitational wave backgrounds from first order phase transitions and sources that induce secondary gravitational waves, as well as opening the microhertz frequency window through precision binary pulsar observations. The synergies between SKAO and other gravitational wave detectors, in particular LISA and the next generation of ground-based observatories, will enable multi-band and multi-messenger studies of massive black hole binaries that are not achievable by any single experiment alone. Furthermore, SKAO's ability to map large-scale structure through intensity mapping and galaxy surveys will allow the anisotropic distribution of gravitational wave sources to be cross-correlated with matter tracers, shedding light on the origin and formation channels of black hole binaries across cosmic history.

\begin{table}[ht]
    \centering
    \begin{tabular}{|l|>{\raggedright\arraybackslash}p{4cm}|>{\raggedright\arraybackslash}p{3cm}|l|}
    \hline
    Chapter & Science target & Observing mode & Configuration \\
    \hline
    \hline
    \cite{Besancon01.2026.SKA} & Memory effects & Pulsar Timing & AA4 \\
    \hline
    \cite{Pasechnik01.2026.SKA} & GWB from FOPT & Pulsar Timing & AA$^*$ \\
    \hline
    \cite{Ragavendra01.2026.SKA} & GWB from Early Universe & Pulsar Timing & AA$^*$ \\
    \cite{Jenkins01.2026.SKA} &  &  &  \\
    \hline
    \cite{Perna01.2026.SKA} & PTA-Astrometry GWB joint-analysis & Pulsar Timing & AA$^*$ \\
    \cite{Capelo01.2026.SKA} &  &  &  \\
    \hline
    \cite{Cruz01.2026.SKA} & GWB dipole & Pulsar Timing & AA4 \\
    \hline
    \cite{Ragavendra01.2026.SKA} & GWB astrophysical anisotropies & Pulsar Timing & AA4 \\
    \hline
    \cite{Bosi01.2026.SKA} & Hz-kHz GW anisotropies & Radio Continuum, Intensity Mapping & AA4 \\
    \hline
    \end{tabular}
\caption{Summary table of the chapters, their main scientific target, the necessary SKAO observing mode, and a tentative minimal SKAO configuration with promising scientific potential.}
\label{tab:summary_table}
\end{table}

Table~\ref{tab:summary_table} contains a summary of the main scientific targets, the relevant SKAO observing mode used to pursue such observations, and a minimal SKAO configuration that presents a promising scientific potential.
However, given the wealth of surprises that GW Science has produced over the past decade, the analysis for all cases should begin already during AA$^*$ to test pipelines and start providing the first upper bounds.

Many of these gravitational wave science cases are closely connected to the pulsars programme, meaning that these targets will be achieved through a coordinated effort across these overlapping groups. Looking ahead, as PTA datasets grow in both the number of monitored pulsars and the precision of timing residuals, and as LISA and third-generation ground-based GW detectors begin operations in the 2030s, the SKAO will be uniquely positioned at the intersection of these observational frontiers. The coming decade promises to be a expansive period for gravitational wave science, and the SKAO will play a key role in extracting new physics across the frequency spectrum.

%%%%%%%%%%%%%%%%%%%%%%%%%%%%%%%%%%%%%%%%%%%%%%%%%%%%%%%%%%%%%%%%%%%%%%%%%%%%%%%%%%%%%%%%%%%%%%%%%%%%%%%%%%%%%%%%%%%%%%%%%%%%%%%%%%%%%%%%%%%%%%%%%%

\section*{Acknowledgments}

NB acknowledges support from the European Union's Horizon Europe research and innovation program under the Marie Sk\l{}odowska-Curie grant agreement no. 101207487 (GWSKY - Mapping the Universe with Gravitational Waves). 

%%%%%%%%%%%%%%%%%%%%%%%%%%%%%%%%%%%%%%%%%%%%%%%%%%%%%%%%%%%%%%%%%%%%%%%%%%%%%%%%%%%%%%%%%%%%%%%%%%%%%%%%%%%%%%%%%%%%%%%%%%%%%%%%%%%%%%%%%%%%%%%%%%

\bibliographystyle{abbrvnat-maxbibnames4}
\bibliography{chapter}

\end{document}